\documentclass[fleqn,twoside]{article}

\usepackage{espcrc2}

\usepackage{graphicx}

\newcommand{\fluxunit}{\mbox{$\mathrm{\,\,GeV\,cm^{-2}\,s^{-1}\,sr^{-1}}$}}
\newcommand{\degr}{\mbox{$^\circ$}}

\title{New results from the Antarctic Muon And Neutrino Detector Array}

\begin{document}

\author{K.~Woschnagg for the AMANDA Collaboration:\\
M.~Ackermann\address[4]{ DESY-Zeuthen, 15735 Zeuthen, Germany\vspace{-2.5mm}},
J.~Ahrens\address[11]{ Institute of Physics, University of Mainz, D-55099 Mainz, Germany\vspace{-2.5mm}},
H.~Albrecht\addressmark[4],
D.~W.~Atlee\address[8]{ Dept.\ of Physics, Pennsylvania State University, University Park, PA 16802, USA\vspace{-2.5mm}},
X.~Bai\address[1]{ Bartol Research Institute, University of Delaware, Newark, DE 19716, USA\vspace{-2.5mm}},
R.~Bay\address[9]{ Dept.\ of Physics, University of California, Berkeley, CA 94720, USA\vspace{-2.5mm}},
M.~Bartelt\address[2]{ Dept.\ of Physics, Bergische Universit\"at Wuppertal, D-42097 Wuppertal, Germany\vspace{-2.5mm}},
S.~W.~Barwick\address[10]{ Dept.\ of Physics and Astronomy, University of California, Irvine, CA 92697, USA\vspace{-2.5mm}},
T.~Becka\addressmark[11],
K.-H.~Becker\addressmark[2],
J.~K.~Becker\addressmark[2],
E.~Bernardini\addressmark[4],
D.~Bertrand\address[3]{ Universit\'e Libre de Bruxelles, Science Faculty CP230, B-1050 Brussels, Belgium\vspace{-2.5mm}},
D.~J.~Boersma\addressmark[4],
S.~B\"oser\addressmark[4],
O.~Botner\address[17]{ Division of High Energy Physics, Uppsala University, S-75121 Uppsala, Sweden\vspace{-2.5mm}},
A.~Bouchta\addressmark[17],
O.~Bouhali\addressmark[3],
J.~Braun\address[15]{ Dept.\ of Physics, University of Wisconsin, Madison, WI 53706, USA\vspace{-2.5mm}},
C.~Burgess\address[18]{ Dept.\ of Physics, Stockholm University, S-10691 Stockholm, Sweden\vspace{-2.5mm}},
T.~Burgess\addressmark[18],
T.~Castermans\address[13]{ University of Mons-Hainaut, 7000 Mons, Belgium\vspace{-2.5mm}},
D.~Chirkin\address[7]{ Lawrence Berkeley National Laboratory, Berkeley, CA 94720, USA\vspace{-2.5mm}},
J.~A.~Coarasa\addressmark[8],
B.~Collin\addressmark[8],
J.~Conrad\addressmark[17],
J.~Cooley\addressmark[15],
D.~F.~Cowen\addressmark[8],
A.~Davour\addressmark[17],
C.~De~Clercq\address[19]{ Vrije Universiteit Brussel, Dienst ELEM, B-1050 Brussels, Belgium\vspace{-2.5mm}},
T.~DeYoung\address[12]{ Dept.\ of Physics, University of Maryland, College Park, MD 20742, USA\vspace{-2.5mm}},
P.~Desiati\addressmark[15],
P.~Ekstr\"om\addressmark[18],
T.~Feser\addressmark[11],
T.~K.~Gaisser\addressmark[1],
R.~Ganugapati\addressmark[15],
H.~Geenen\addressmark[2],
L.~Gerhardt\addressmark[10],
A.~Goldschmidt\addressmark[7],
A.~Gro\ss\addressmark[2],
A.~Hallgren\addressmark[17],
F.~Halzen\addressmark[15],
K.~Hanson\addressmark[15],
D.~Hardtke\addressmark[9],
R.~Hardtke\addressmark[15],
T.~Harenberg\addressmark[2],
T.~Hauschildt\addressmark[4],
K.~Helbing\addressmark[7],
M.~Hellwig\addressmark[11],
P.~Herquet\addressmark[13],
G.~C.~Hill\addressmark[15],
J.~Hodges\addressmark[15],
D.~Hubert\addressmark[19],
B.~Hughey\addressmark[15],
P.~O.~Hulth\addressmark[18],
K.~Hultqvist\addressmark[18],
S.~Hundertmark\addressmark[18],
J.~Jacobsen\addressmark[7],
K.-H.~Kampert\addressmark[2],
A.~Karle\addressmark[15],
J.~Kelley\addressmark[15],
M.~Kestel\addressmark[8],
L.~K\"opke\addressmark[11],
M.~Kowalski\addressmark[4],
M.~Krasberg\addressmark[15],
K.~Kuehn\addressmark[10],
H.~Leich\addressmark[4],
M.~Leuthold\addressmark[4],
J.~Lundberg\addressmark[17],
J.~Madsen\address[16]{ Physics Dept., University of Wisconsin, River Falls, WI 54022, USA\vspace{-2.5mm}},
K.~Mandli\addressmark[15],
P.~Marciniewski\addressmark[17],
H.~S.~Matis\addressmark[7],
C.~P.~McParland\addressmark[7],
T.~Messarius\addressmark[2],
Y.~Minaeva\addressmark[18],
P.~Mio\v{c}inovi\'{c}\addressmark[9],
R.~Morse\addressmark[15],
K.~M\"unich\addressmark[2],
R.~Nahnhauer\addressmark[4],
J.~W.~Nam\addressmark[10],
T.~Neunh\"offer\addressmark[11],
P.~Niessen\addressmark[1],
D.~R.~Nygren\addressmark[7],
H.~\"Ogelman\addressmark[15],
Ph.~Olbrechts\addressmark[19],
C.~P\'erez~de~los~Heros\addressmark[17],
A.~C.~Pohl\address[6]{ Dept.\ of Technology, Kalmar University, S-39182 Kalmar, Sweden\vspace{-2.5mm}},
R.~Porrata\addressmark[9],
P.~B.~Price\addressmark[9],
G.~T.~Przybylski\addressmark[7],
K.~Rawlins\addressmark[15],
E.~Resconi\addressmark[4],
W.~Rhode\addressmark[2],
M.~Ribordy\addressmark[13],
S.~Richter\addressmark[15],
J.~Rodr\'{i}guez~Martino\addressmark[18],
H.-G.~Sander\addressmark[11],
K.~Schinarakis\addressmark[2],
S.~Schlenstedt\addressmark[4],
D.~Schneider\addressmark[15],
R.~Schwarz\addressmark[15],
S.~H.~Seo\addressmark[8],
A.~Silvestri\addressmark[10],
M.~Solarz\addressmark[9],
G.~M.~Spiczak\addressmark[16],
C.~Spiering\addressmark[4],
M.~Stamatikos\addressmark[15],
D.~Steele\addressmark[15],
P.~Steffen\addressmark[4],
R.~G.~Stokstad\addressmark[7],
K.-H.~Sulanke\addressmark[4],
I.~Taboada\address[14]{ Dept.\ of Physics, Universidad Sim\'on Bol\'{\i}var, Caracas, 1080, Venezuela\vspace{-2.5mm}},
O.~Tarasova\addressmark[4],
L.~Thollander\addressmark[18],
S.~Tilav\addressmark[1],
J.~Vandenbroucke\addressmark[9],
L.~C.~Voicu\addressmark[8],
W.~Wagner\addressmark[2],
C.~Walck\addressmark[18],
M.~Walter\addressmark[4],
Y.~R.~Wang\addressmark[15],
C.~H.~Wiebusch\addressmark[2],
R.~Wischnewski\addressmark[4],
H.~Wissing\addressmark[4],
K.~Woschnagg\addressmark[9],
G.~Yodh\addressmark[10]
}

\begin{abstract}
We present recent results from the Antarctic Muon And Neutrino
Detector Array (AMANDA) on searches for high-energy neutrinos
of extraterrestrial origin.
We have searched for a diffuse flux of neutrinos, neutrino point
sources and neutrinos from GRBs and from WIMP annihilations in
the Sun or the center of the Earth.
We also present a preliminary result on the first energy spectrum
above a few TeV for atmospheric neutrinos.
\vspace{1pc}
\end{abstract}

\maketitle

\section{INTRODUCTION}

The existence of high-energy cosmic neutrinos is suggested by the
observation of high-energy cosmic rays and gamma rays.
Observation of cosmic neutrinos could shed light on the production
and acceleration mechanisms of cosmic rays, which are not understood
for energies above the ``knee'' at $10^{15}$~eV.
Neutrinos with energies in the TeV range and higher may be produced
by a variety of sources.
Candidate cosmic accelerators include supernova remnants, the accretion
disk and jets of Active Galactic Nuclei (AGN), and the violent processes
behind Gamma Ray Bursts (GRB).
In these environments, neutrinos are expected to be produced in the decays
of pions created through proton-proton or proton-photon collisions.
The AMANDA detector was built to explore the high-energy universe in
neutrinos, using the advantages of neutrinos as cosmic messengers.
In January 2005, construction will begin on IceCube~\cite{OlgaParis},
the km$^3$-sized successor to AMANDA.

\section{THE AMANDA DETECTOR}

The AMANDA detector\footnote{The full 19-string array, named
AMANDA-II, started taking data in 2000. An earlier 10-string stage
(comprising the inner 10 strings), called AMANDA-B10, was taking
data in the period 1997--1999.} \cite{NaturePaper} consists of 677
optical modules arranged along 19 vertical strings buried deep in
the glacial ice at the South Pole, mainly at depths between 1500
and 2000 m. Each module consists of a photomultiplier tube (PMT)
housed in a spherical glass pressure vessel. PMT pulses are
transmitted to the data acquisition electronics at the surface via
coaxial cables (inner 4 strings), twisted pair cables (6 strings)
or optical fibers (outer 9 strings). The geometric outline of the
array is a cylinder which is 500 m high and with a radius of 100
m. The typical vertical spacing between modules is 10--20 m, and
the horizontal spacing between strings 30--50 m.

The optical modules record Cherenkov light generated by secondary charged
leptons (e, $\mu$, $\tau$) created in neutrino interactions near the detector.
Events are reconstructed by maximizing the likelihood that the timing
pattern of the recorded light is produced by a hypothetical track or cascade
\cite{Areco}.
The angular resolution is between 1.5\degr\ and 2.5\degr\ for muon tracks,
depending on declination, and $\sim$30$\degr$ for cascades,
the difference reflecting the fact that muon tracks yield a long lever arm
whereas cascades produce more spherical light patterns.
On the other hand, the energy resolution, which is correlated to the amount of
detected light, is better for cascades, 0.15 in $\log(E)$, than for
muon tracks, 0.4 in $\log(E)$.

\begin{figure}[p]
\includegraphics[width=\columnwidth]{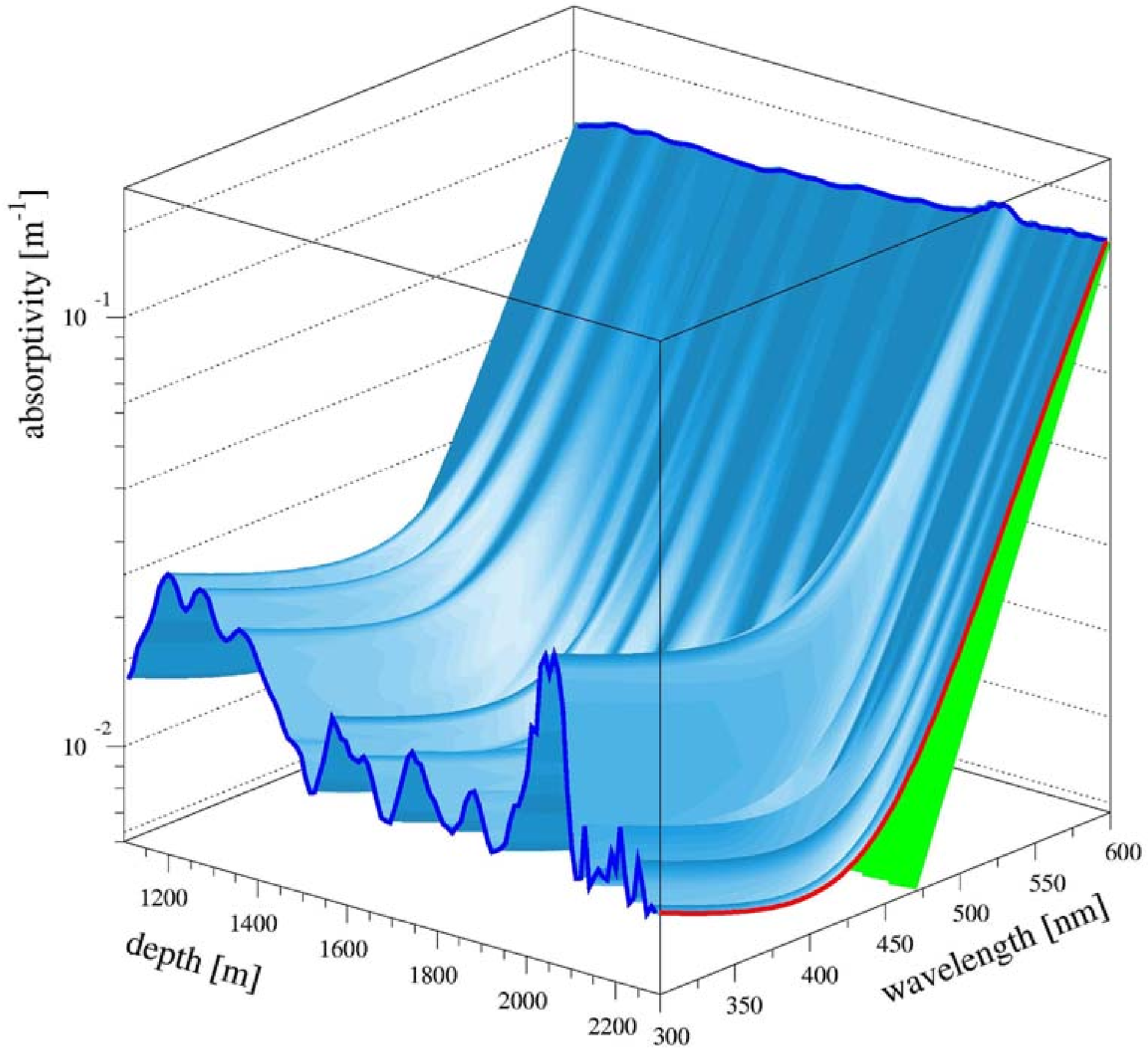}\\
\includegraphics[width=\columnwidth]{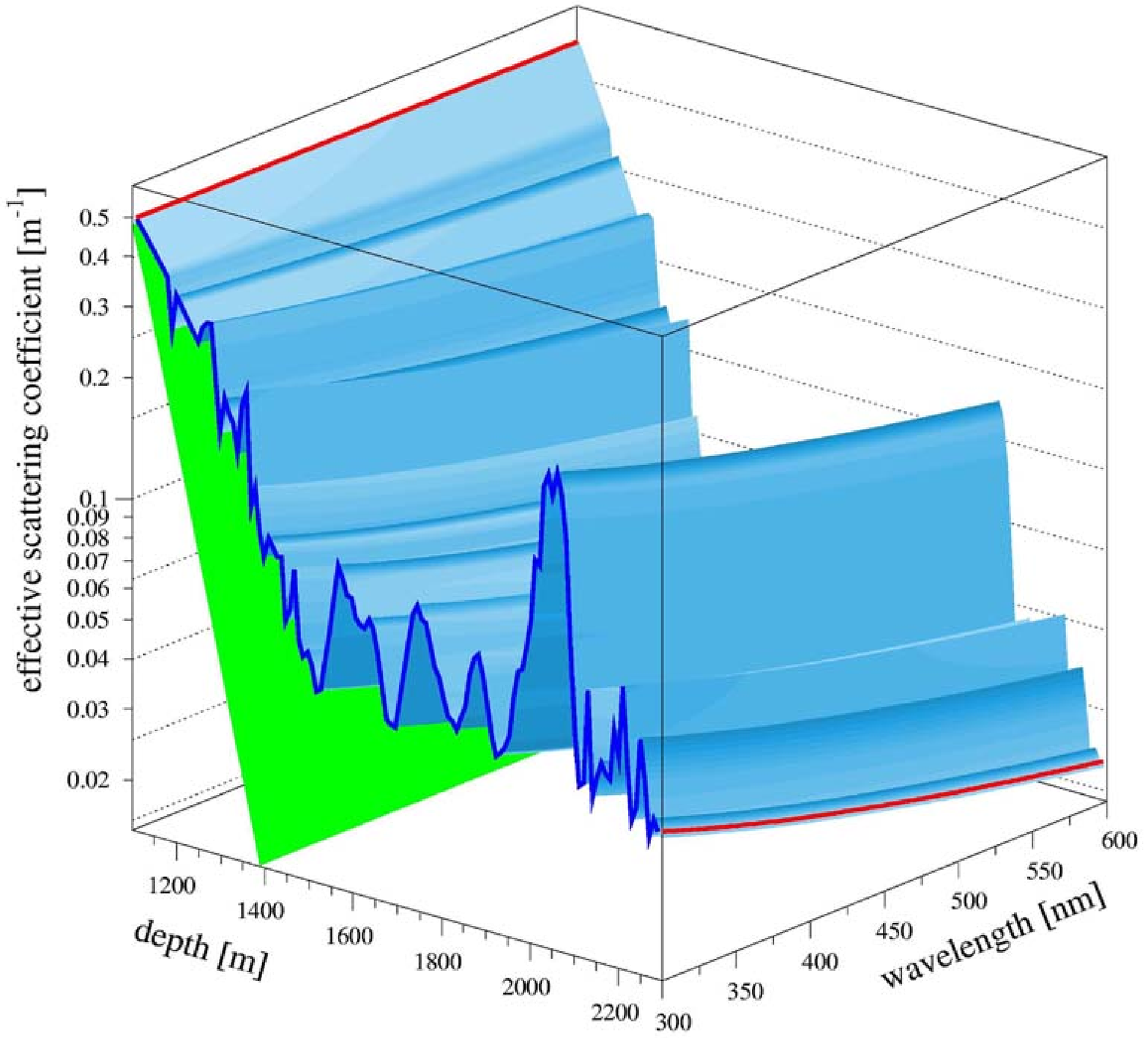}
\caption{Optical properties of deep South Pole ice:
absorptivity (top) and scattering coefficient (bottom) as function of depth
and wavelength. The green (partially obscured) tilted planes show the contribution
from pure ice to absorption and from air bubbles to scattering, respectively.
If these contributions are subtracted, the optical properties vary with the
concentration of insoluble dust, which tracks climatological variations in the past
\cite{Vostok,DomeC,DomeFuji}.}
\label{fig:iceprop}
\end{figure}

Using calibration light sources deployed with the strings and a YAG laser
at the surface connected to diffusive balls in the ice via optical fibers,
we have mapped~\cite{icepaper} the optical properties of the ice over the
full relevant wavelength- and depth range (figure~\ref{fig:iceprop}).
The glacial ice is extremely transparent for Cherenkov wavelengths
near the peak sensitivity of the modules:
at 400 nm, the average absorption length is 110 m and the average effective
scattering length is 20 m.
Below a depth of 1500 m, both scattering and absorption are dominated by
dust, and the optical properties vary with dust concentration.
The depth profile is in good agreement with variations of dust concentration
measured in ice cores from other Antarctic sites \cite{Vostok,DomeC,DomeFuji}.
These dust layers reflect past variations in climate.
Implementation of the detailed knowledge of ice properties into our detector
simulation and reconstruction tools reduces systematic uncertainties
and improves track and cascade reconstruction.

In the 2003/04 field season, the data acquisition system was upgraded
with Transient Waveform Recorders on all channels, digitizing the PMT pulses
in the electronics on the surface.
Waveform digitization will increase the effective dynamic range of individual
channels by about a factor 100 and will lead to an improvement in energy
reconstruction, especially at high energies.

\section{PHYSICS TOPICS AND ANALYSIS STRATEGIES}

AMANDA is used to explore a variety of physics topics, ranging from
astrophysics to particle physics, over a wide range of energies.
At the lower energy end, in the MeV range, AMANDA is sensitive to
fluxes of antineutrinos from supernovae. For higher energies, GeV to TeV,
the detector is used to study atmospheric neutrinos and to conduct
indirect dark matter searches. In the energy range for which AMANDA has been
primarily optimized, TeV to PeV, the aim is to use neutrinos to study AGN and GRBs,
looking both for a diffuse flux and for point sources of high-energy neutrinos.
Using special analysis techniques, the array is also sensitive to the
ultra-high energies in the PeV to EeV range.

For most analysis channels, AMANDA uses the Earth as a filter and
looks down for up-going neutrinos.
The main classes of background are up-going atmospheric neutrinos and
down-going atmospheric muons that are misreconstructed as up-going.
Since AMANDA is located at the South Pole, an up-going event will
have originated in the Northern sky.

We present all flux limits following the ordering scheme in \cite{FC}
and include systematic uncertainties in the limit calculations
according to the method derived in \cite{Conrad}.
The main sources of systematic uncertainty in the analyses presented here
are the modelling of muon propagation and of optical ice properties
in the detector simulation, adding up to roughly 25\% uncertainty.

The AMANDA collaboration adheres strictly to a policy of performing all
analyses in a ``blind'' manner to ensure statistical purity of the results.
In practice, this means that selection criteria are optimized either on a
sub-sample of the data set which is then excluded from the analysis yielding the
final result, or on a time-scrambled data set which is only unscrambled
after the selection criteria have been optimized and finalized.

\section{ATMOSPHERIC NEUTRINOS}

\begin{figure}[htb]
\begin{center}
\includegraphics[width=\columnwidth]{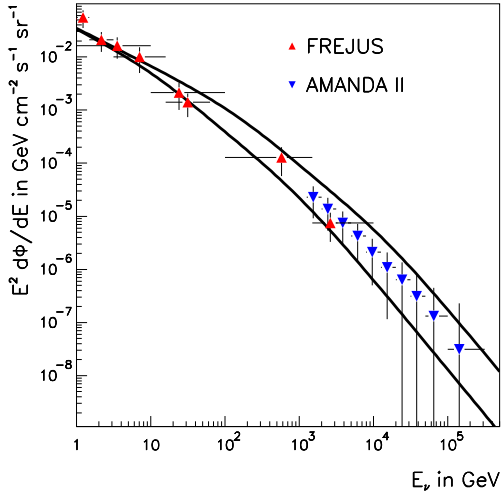}
\caption{Atmospheric neutrino energy spectrum (preliminary) from regularized unfolding
of AMANDA data, compared to the Frejus spectrum \cite{Frejus} at lower energies. The two
solid curves indicate model predictions \cite{Volkova} for the horizontal
(upper) and vertical (lower) flux.}
\label{fig:unfoflux}
\end{center}
\end{figure}

Neutrinos, and to some extent muons, created by cosmic ray interactions
in the atmosphere constitute the main background in most analysis channels,
but also serve as a test beam.
Using a neural net energy reconstruction, trained on a full detector and
physics simulation, followed by regularized unfolding, we measure a preliminary
energy spectrum for up-going neutrinos with AMANDA-II data from 2000 (figure~\ref{fig:unfoflux}).
This is the first atmospheric neutrino spectrum above a few TeV, and it extends
up to 300 TeV.

\section{SEARCHES FOR A DIFFUSE FLUX OF COSMIC NEUTRINOS}

The ultimate goal of AMANDA is to find and study the properties of cosmic
sources of high-energy neutrinos.
Should individual sources be too weak to produce an unambiguous directional
signal in the array, the integrated neutrino flux from all sources could
still produce a detectable diffuse signal.
We have searched several years of data for such a diffuse signal using
complementary techniques in different energy regimes.

\subsection{Atmospheric neutrino spectrum}

The atmospheric neutrino spectrum (fig.~\ref{fig:unfoflux}) was used to set an upper limit
on a diffuse $E^{-2}$ flux of extraterrestrial muon neutrinos for the energy range covered by
the highest bin, 100--300 TeV, by calculating the maximal non-atmospheric contribution
to the flux in the bin given its statistical uncertainty.
However, the bins in the unfolded spectrum are correlated and the uncertainty
in the last bin can not a priori be assumed to be Poissonian. The statistics
in the bin were therefore determined with a Monte Carlo technique used to
construct confidence belts following the definition by Feldman and Cousins~\cite{FC}.
Given the unfolded number of experimental events in the bin (a fractional number),
a preliminary 90\% C.L.\ upper limit of
\begin{equation}
E^2 \Phi_{\nu_{\mu}}(E) < 2.6 \times 10^{-7} \fluxunit
\end{equation}
was derived for $100~\mathrm{TeV} < E_{\nu} < 300~\mathrm{TeV}$,
which includes 33\% systematic uncertainties.

\subsection{Cascades}

In the cascade channel, AMANDA has essentially $4\pi$ coverage, and
is sensitive to all three neutrino flavors.
The 2000 data sample, corresponding to 197 days livetime, was searched
for cascade events.
Event selection was based on topology and energy, and optimized
to maximize the sensitivity to an $E^{-2}$ signal spectrum.
After final cuts one event remains, with an expected background
of $0.90^{+0.69}_{-0.43}$ from atmospheric muons and $0.06^{+0.09}_{-0.04}$
from atmospheric neutrinos.
Not having observed an excess over background, we calculate a limit
on a signal flux.
The 90\% C.L.\ limit on a diffuse flux of neutrinos of all flavors
for neutrino energies between 50 TeV and 5 PeV, assuming full flavor
mixing so that the neutrino flavor ratios are 1:1:1 at the detector,
is
\begin{equation}
E^2 \Phi_{\nu}(E) < 8.6 \times 10^{-7} \fluxunit.
\end{equation}
Since the energy range for this analysis contains the energy of the
Glashow resonance (6.3 PeV) the above limit translates to
\begin{equation}
E^2 \Phi_{\bar\nu_{\mathrm{e}}}(6.3\,\mathrm{PeV}) < 2 \times 10^{-6} \fluxunit.
\end{equation}
These limits \cite{AIIcasc} obtained with one year (2000) of AMANDA-II data
are roughly a factor 10 lower than the limits from similar searches performed
with AMANDA-B10 data from 1997 \cite{B10casc} and 1999 \cite{AIIcasc}.

\subsection{Ultra High Energy neutrinos}

At ultra-high energies (UHE), above 1 PeV, the Earth is opaque
to electron- and muon-neutrinos. Tau neutrinos with such initial energies might
penetrate the Earth through regeneration~\cite{tauregen}, in which the $\tau$
produced in a charged-current $\nu_{\tau}$ interaction decays back into $\nu_{\tau}$,
but they will emerge with much lower energies.
The search for extraterrestrial UHE neutrinos is therefore concentrated
on events close to the horizon and even from above.
The latter is possible since the atmospheric muon background is low
at these high energies due to the steeply falling spectrum.
Our search for UHE events in 1997 AMANDA-B10 data (131 days of livetime) relies on
parameters that are sensitive to the expected characteristics of an UHE signal:
bright events, long tracks (for muons), low fraction of single photoelectron hits.
A neural net was trained to optimize the sensitivity to an $E^{-2}$ neutrino signal
in data dominated by atmospheric neutrino background.

After final selection, 5 data events remain, with $4.6$ ($\pm 36\%$) expected
background. Thus, no excess above background is observed and we derive \cite{UHE97}
a 90\% C.L.\ limit on an $E^{-2}$ flux of neutrinos of all flavors,
assuming a 1:1:1 flavor ratio at Earth, for energies between 1 PeV and 3 EeV, of
\begin{equation}
E^2 \Phi_{\nu}(E) < 0.99 \times 10^{-6} \fluxunit.
\end{equation}
A similar analysis of AMANDA-II data from 2000 is under way.
However, the bright UHE events also saturate the larger array, so a substantial
gain in sensitivity will mainly be due to the additional exposure time and
improved selection algorithms.

\subsection{Summary of diffuse searches}

Using different analysis techniques, AMANDA has set limits on the diffuse
flux of neutrinos with extraterrestrial origin for neutrino energies
from 6 TeV \cite{Diff97} up to a few EeV.
With the exception of the limit from the unfolded atmospheric spectrum,
which can be seen as a quasi-differential limit, the limits are on the
integrated flux over the energy range which contains 90\% of the signal.
Our limits exclude, at 90\% C.L., some models \cite{SSmodel,Nmodel}
predicting diffuse neutrino fluxes.

\section{POINT SOURCE SEARCHES}

\begin{figure}[htb]
\begin{center}
\includegraphics[width=\columnwidth]{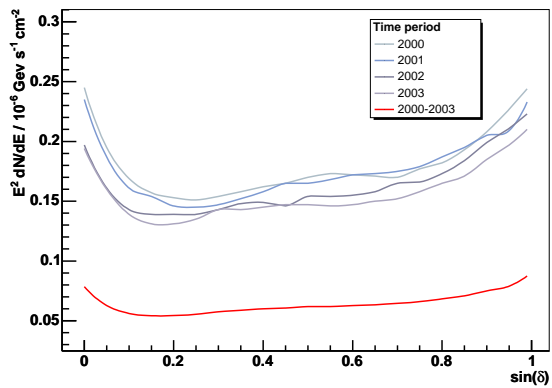}
\caption{AMANDA-II sensitivity for an $E^{-2}$ flux spectrum as function of declination.}
\label{fig:sensit}
\end{center}
\end{figure}

Searches for neutrino point sources require good pointing resolution and are
thus restricted to the $\nu_\mu$ channel.
We have searched AMANDA-II data from 2000--2003 (807 days livetime)
for a point source signal.
Events were selected to maximize the model rejection potential~\cite{MRFpaper}
for an $E^{-2}$ neutrino spectrum convoluted with the background spectra due
to atmospheric neutrinos and misreconstructed atmospheric muons.
The selection criteria were optimized for the combined 4-year data set in each
declination band separately, since the geometry of the detector array introduces
declination-dependent efficiencies.
The {\it sensitivity} of the analysis, defined as the average upper limit one would expect
to set on a non-atmospheric neutrino flux if no signal is detected, is shown in
figure~\ref{fig:sensit} for a hypothetical $E^{-2}$ signal spectrum.

\begin{figure}[htb]
\begin{center}
Reconstructed sky coordinates\\[1mm]
\includegraphics[width=\columnwidth,bb=30 20 540 210]{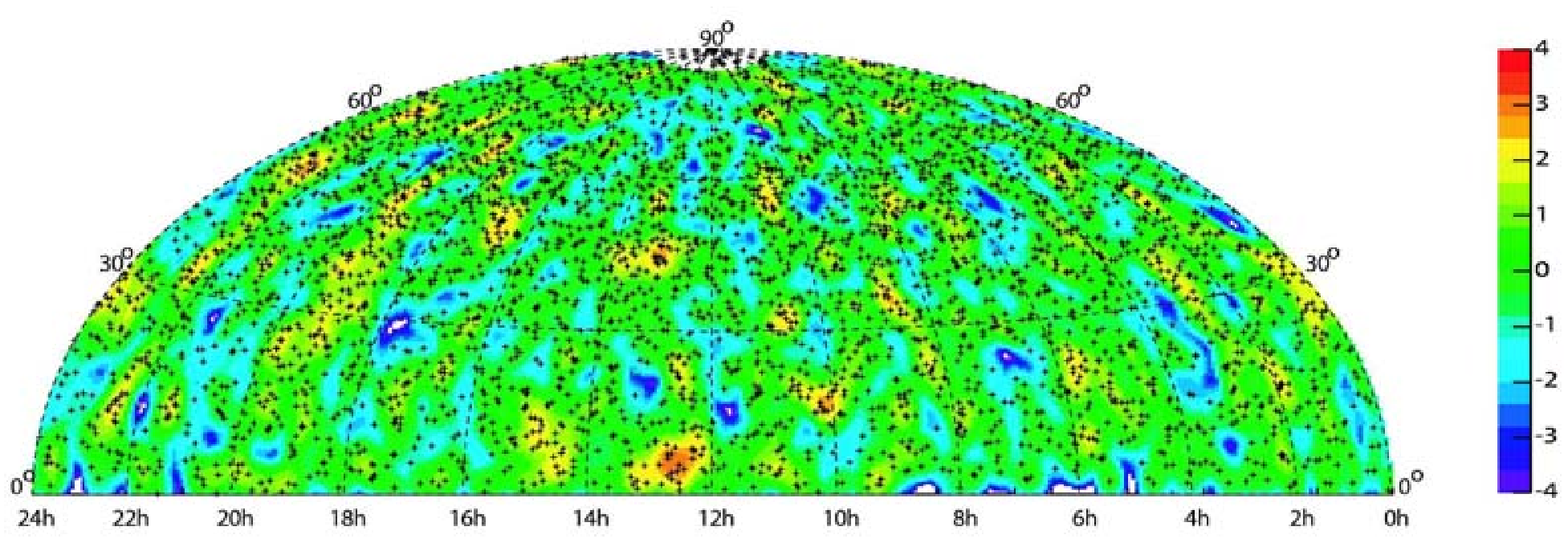}\\[5mm]
Scrambled in right ascension\\[2mm]
\includegraphics[width=\columnwidth,bb=30 25 520 205]{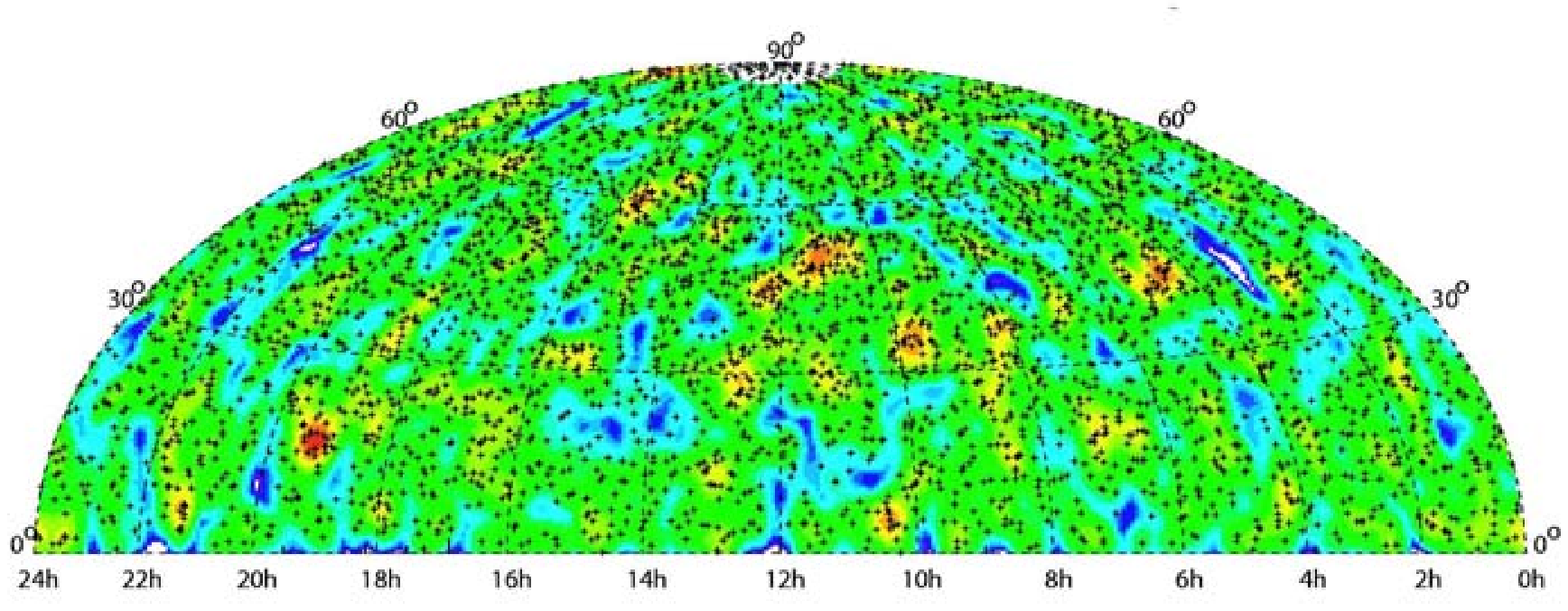}
\caption{Significance map (top) constructed from 3369 events in the final sample from a
point source search with AMANDA-II data from 2000--2003. The points show the reconstructed
sky positions (declination and right ascension) of the neutrino candidates.
The color scale indicates the significance (in $\sigma$).
The lower panel shows an example of a significance map based on the same events,
but with randomized right ascension coordinates.}
\label{fig:skyplots}
\end{center}
\end{figure}

The final sample of 3369 neutrino candidates (with 3438 expected atmospheric neutrinos)
was searched for point sources with two methods.
In the first, the sky is divided into a (repeatedly shifted) fine-meshed grid of
overlapping bins which are tested for a statistically significant excess over the
background expectation (estimated from all other bins in the same declination band).
This search yielded no evidence for extraterrestrial point sources.
The second method is an unbinned search, in which the sky locations of the events and
their uncertainties from reconstruction are used to construct a sky map of significance
in terms of fluctuation (in $\sigma$) over background (figure~\ref{fig:skyplots}).
This map displays only one potential hot spot (above 3$\sigma$), which is well within
the expectation from a random event distribution.
For comparison, the same significance map was constructed after randomizing
the right ascension for all events, thus simulating a truly random distribution
(lower panel in the figure).
This scrambled map is statistically indistinguishable from the real (upper) map.
A full statistical analysis of many such scrambled maps proves that the sky map is
fully compatible with a distribution expected from an atmospheric neutrino sample.
We thus see no evidence for point sources with an $E^{-2}$ energy spectrum
based on the first four years of AMANDA-II data.
This preliminary result complements previously published results from point
source searches with the AMANDA-B10 detector \cite{PS1997} and the first year
of AMANDA-II data \cite{PS2000}.

\section{SEARCH FOR NEUTRINOS FROM GRBs}

A special case of point source analysis is the search for neutrinos
coincident with gamma ray bursts (GRBs) detected by satellite-borne
detectors.
For this search, the timing of the neutrino event serves as an additional
selection handle which significantly reduces background.

We have used the GRB sample collected by the BATSE instrument on board the CGRO satellite.
The AMANDA and BATSE data taking periods were overlapping between 1997,
when AMANDA-B10 became operational, and 2000, when CGRO was decommissioned.
In total, we have analyzed a sample containing 312 bursts triggered by BATSE
from this period.
For each of these bursts, AMANDA data was searched for an excess over background
of events in a 10 min window around the GRB time (here defined as the start of $T_{90}$).
The background was estimated by averaging over events in the on-source spatial bin within
$\pm$1 hour of the burst (excluding the 10 min signal window).

No neutrino event was observed in coincidence with any of the bursts.
Assuming a broken power-law energy spectrum as proposed by Waxmann and Bahcall
\cite{WB}, with $E_{\mathrm{break}}=100$ TeV and $\Gamma_{\mathrm{bulk}}=300$,
we obtain a 90\% C.L.\ upper limit on the expected neutrino flux at the Earth of
\begin{equation}
E^2 \Phi_{\nu}(E) < 4 \times 10^{-8} \fluxunit.
\end{equation}
This is approximately a factor 15 above the Waxmann-Bahcall flux prediction.

Work is under way to include other classes of bursts in the analysis.
A class of bursts that did not trigger the BATSE detector but were found by
a later off-line analysis of archived data \cite{nonBATSE} comprises 26 events
in the Northern sky during the up-time of AMANDA in 2000.
Since 2000, the only source of GRB detection is the Third Interplanetary Network (IPN3),
a group of spacecraft equipped with gamma-ray burst detectors which uses triangulation
to spatially locate the bursts.
IPN-triggered bursts will also be included in future GRB-neutrino searches with
AMANDA.

\section{DARK MATTER SEARCHES}

\begin{figure}[p]
\begin{center}
\includegraphics[width=\columnwidth,bb= 0 0 400 400]{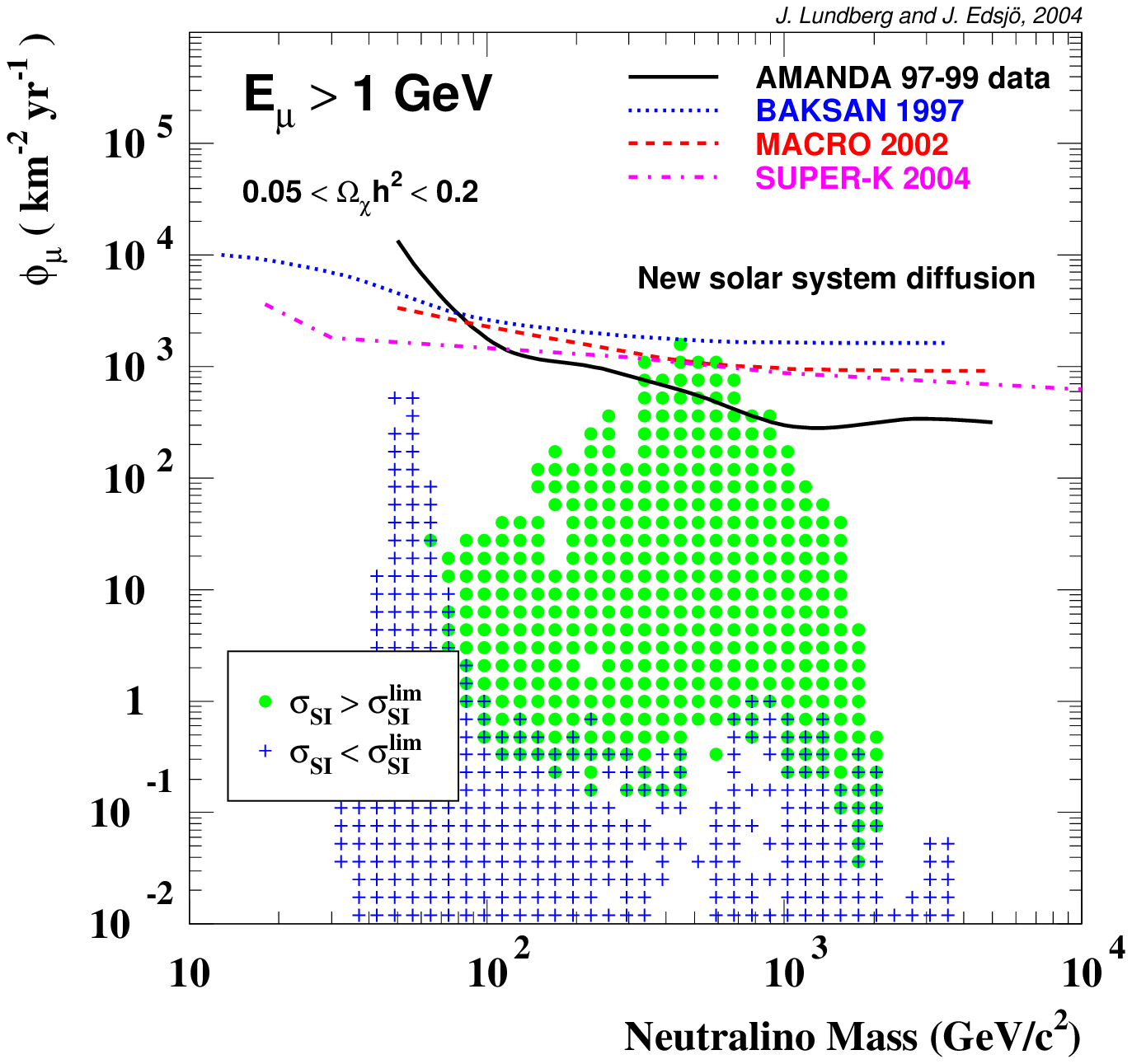}\\
\includegraphics[width=\columnwidth,bb= 0 0 400 400]{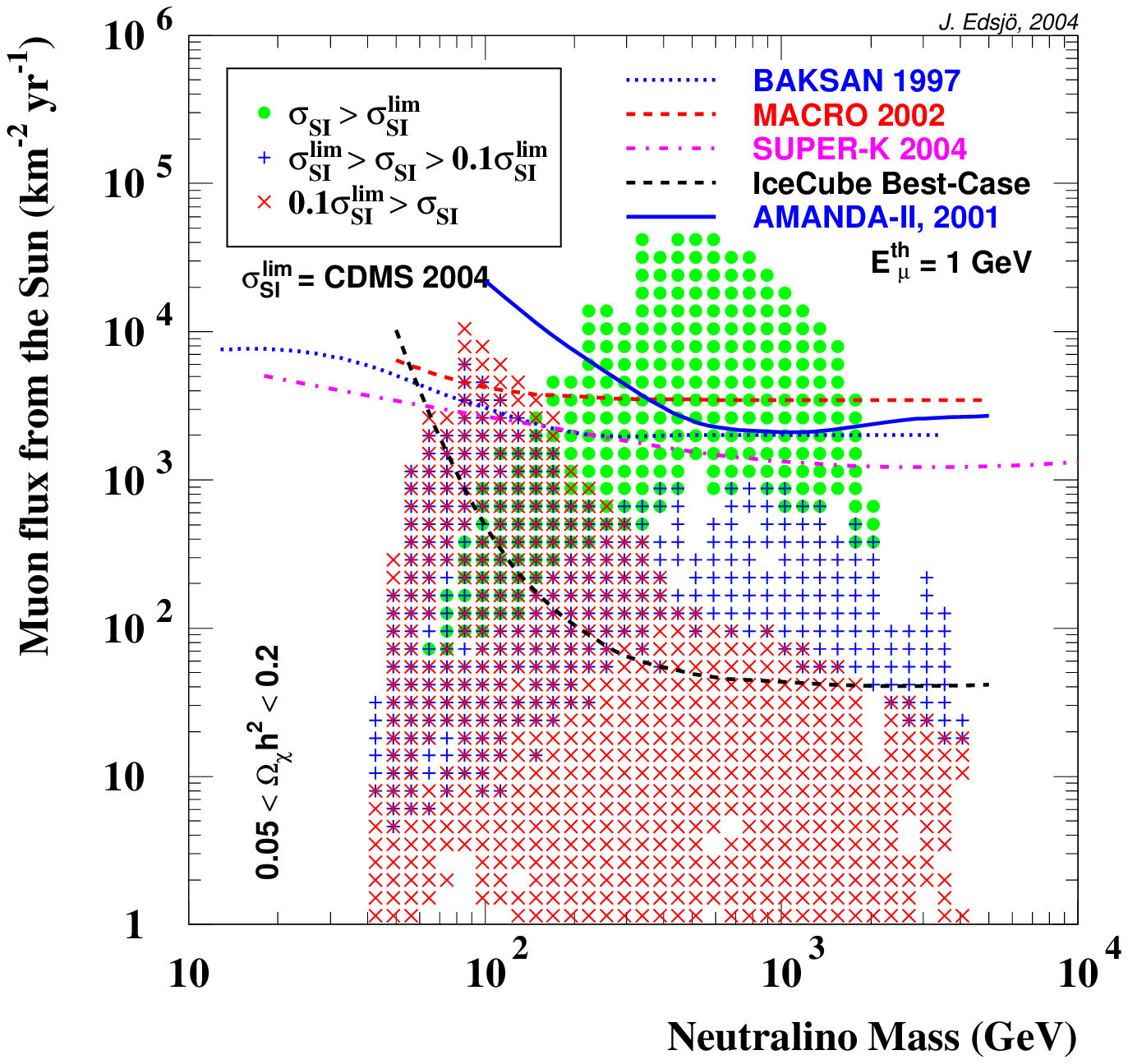}
\caption{Preliminary limits on the muon flux due to neutrinos from
neutralino annihilations in the center of the Earth (top) and the Sun (bottom). The
colored symbols correspond to model predictions \cite{JoakimParis} within the allowed
parameter space of the MSSM.
The green models are disfavored by direct searches with CDMS II~\cite{CDMS2004}.}
\label{fig:wimplimits}
\end{center}
\end{figure}

Particle physics provides an interesting candidate for non-baryonic dark matter
in the Weakly Interacting Massive Particle (WIMP).
In particular, the Minimal Supersymmetric extension of the Standard Model
(MSSM) provides a promising WIMP candidate in the neutralino, which could be the
lightest supersymmetric particle.
Neutralinos can be gravitationally trapped in massive bodies, and can then via
annihilations and the decay of the resulting particles produce neutrinos.
AMANDA can therefore perform indirect dark matter searches by looking
for fluxes of neutrinos from the center of the Earth or the Sun.

For the former, we present a preliminary update to our published limits
obtained with one year of 10-string data \cite{B10wimps}.
We have looked for vertically up-going tracks in AMANDA-B10 data from
1997 to 1999, corresponding to a total livetime of 422 days.
No WIMP signal was found and a 90\% C.L.\ upper limit on the muon flux
from the center of the Earth was set for neutralino masses between 50 GeV
and 5 TeV (figure~\ref{fig:wimplimits}, upper panel).

Due to its larger mass (resulting in a deeper gravitational well) and
a higher capture rate due to additional spin-dependent processes,
the Sun can also be used for WIMP searches despite its much larger distance
from the detector.
Although the Sun is maximally 23\degr\ below the horizon at the South Pole,
AMANDA-II can be used for a WIMP search thanks to its improved reconstruction
capabilities for horizontal tracks.
Analysis of 2001 data (0.39 years of livetime) yielded no WIMP signal.
The preliminary upper limit on the muon flux from the Sun is compared
to MSSM predictions~\cite{JoakimParis} in figure~\ref{fig:wimplimits} (lower panel).
For heavier neutralino masses, the limit obtained with less than one year of AMANDA-II
data is already competitive with limits from indirect searches with detectors
that have several years of integrated livetime.
The green points in figure~\ref{fig:wimplimits} correspond to models that are
disfavored by direct searches \cite{CDMS2004}, which appear to set more severe
restrictions on the allowed parameter space than indirect searches.
However, it should be noted that the two methods are complementary in that they
(a) probe the WIMP distribution in the solar system at different epochs
and (b) are sensitive to different parts of the velocity distribution.

\section{SUPERNOVA DETECTION}

Since 2003 the AMANDA supernova system includes all AMANDA-II channels.
Recent upgrades of the online analysis software have improved the supernova
detection capabilities such that AMANDA-II can detect 90\% of supernovae within
9.4 kpc with less than 15 fakes per year.
This is sufficiently robust for AMANDA to now contribute to the SuperNova Early Warning System
(SNEWS) with neutrino detectors in the Northern hemisphere.

\section*{ACKNOWLEDGEMENTS}

{\small
We acknowledge the support of the following agencies: National
Science Foundation -- Office of Polar Programs, National Science
Foundation -- Physics Division, University of Wisconsin Alumni
Research Foundation, Department of Energy and National Energy
Research Scientific Computing Center (supported by the Office of
Energy Research of the Department of Energy), UC-Irvine ANEAS
Supercomputer Facility, USA; Swedish Research Council, Swedish
Polar Research Secretariat and Knut and Alice Wallenberg
Foundation, Sweden; German Ministry for Education and Research,
Deutsche Forschungsgemeinschaft (DFG), Germany; Fund for
Scientific Research (FNRS-FWO), Flanderns Institute to encourage
Scientific and Technological Research in Industry (IWT) and
Belgian Federal Office for Scientific, Technical and Cultural
affairs (OSTC), Belgium; Fundaci$\acute{\mbox{o}}$n Venezolana de
Promoci$\acute{\mbox{o}}$n al Investigador (FVPI), Venezuela;
D.F.C.\ acknowledges the support of the NSF CAREER program;
E.R.\ acknowledges the support of the Marie-Curie fellowship program of
the European Union;
M.R.\ acknowledges the support of the Swiss National Science Foundation.
}

\end{document}